\documentclass{article}
\usepackage{hyperref}

\usepackage{amsmath,epsfig}
\usepackage{amsfonts,amssymb}

\newcommand{\ba}{\left( \begin{array}}
\newcommand{\ea}{\end{array} \right)}
\newcommand{\bq}{\begin{eqnarray*}}
\newcommand{\eq}{\end{eqnarray*}}
\newcommand{\bqn}{\begin{eqnarray}}
\newcommand{\eqn}{\end{eqnarray}}

\usepackage{hyperref}
\hypersetup{
    colorlinks=true,%
    citecolor=blue,%
    filecolor=blue,%
    linkcolor=red,%
    urlcolor=blue,}
\newcommand{\red}[1]{{\color{red} #1}}
\begin{document}

\title{Statistical Challenges of Big Brain Network Data}
\author{Moo K. Chung
\thanks{This research was partially supported by NIH Brain Initiative Grant R01 EB022856.}\\
University of Wisconsin-Madison, USA\\
\red{\tt mkchung@wisc.edu}}

\maketitle

\begin{abstract}
We explore the main characteristics of big brain network data that offer unique statistical challenges. The brain networks are biologically expected to be both sparse and hierarchical.  Such unique  characterizations put specific topological constraints onto statistical approaches and models we can use effectively. We explore the limitations of the current models used in the field and offer alternative approaches and explain new challenges. 
\end{abstract}

\section{Introduction}

Wikipedia defines {\em big data} as data sets that are so large or complex that traditional data processing application software is inadequate to deal with them (\href{http://en.wikipedia.org/wiki/Big_data}{\tt en.wikipedia.org/wiki/Big\_data}). Big data is not just about the size of the data although that is the main obstacle of using traditional statistical approaches. Big data usually include data sets with sizes beyond the ability of standard software tools to process and analyze within a reasonable time limit. Even 100MB of data can be big if existing computing resources can only handle 1MB of data at a time. Thus, the size of the data is a {\em relative} quantity respect to the available computing resources.

If we pick any article in big data literature these days, chances are that we often encounter hardware solutions  to solving big data problems. They often suggest increasing more central processing units (CPU) or graphical processing units (GPU) and emphasize the need for cluster or parallel computing. For instance, \cite{boubela.2016} suggests to use parallel computing  as a way to compute large-scale Pearson correlation coefficients for 390GB of data in the Human Connectome Project (HCP) but did not suggest any other simpler algorithmic approaches that can be implemented in a limited computing resource environment. Simply adding more hardware is not necessarily an effective but costly strategy for big data. Such hardware approaches often do not provide a venue for more interesting statistical problems. Further, the access to fast computational resources is not necessarily given to everyone. Many biological laboratories still do not have technical expertise of using cluster or parallel computing. Therefore, it is often necessary to develop more algorithmic and statistical approaches in addressing big data at least for biological sciences. 

In this paper, we focus on the statistical challenges of big data in brain imaging and networks that are somewhat different from more traditional big data problems.

\section{Large-scale brain imaging data}
Many big datasets introduce unique computational and statistical challenges that include scalability, storage bottleneck, data representation visualization, and computation mostly related to sample sizes \cite{fan.2014}. However, the challenges in big brain imaging datasets such as HCP and  Alzheimer's Disease Neuroimaging Initiative (ADNI ; \href{http://adni.loni.usc.edu}{\tt adni.loni.usc.edu}) are slightly different. There are substantially more number of voxels ($p$) per image than the number of images ($n$) in the datasets. Even at 3mm low resolution, functional magnetic resonance images (fMRI) has more than 25000 voxels \cite{chung.2017.IPMI}. Unless the dataset consists of more than 25000 images, brain imaging is often the problem of {\em small-$n$ large-$p$}, which is different from the usual big data setting where $n$ is often big. HCP and ADNI have $n$ in the range of a thousands, far smaller than the number of voxels. 

Traditionally, numerical accuracy has been less of concerns in brain imaging particularly due to spatial and temporal smoothing often done in images to smooth out various image processing artifacts and physiological noises. Due to the increased sample size and the central limit theorem, which is further reinforced by smoothing, the statistical distribution of the data might become less of a concern in big imaging data \cite{salmond.2002}.

In the traditional mass univariate approaches \cite{chung.2015.TMI,worsley.1992}, where statistical inference is done at each voxel, the problem of {\em small-$n$ large-$p$} is not critical. Further, spatial smoothing has the effect of reducing the number of {\em resolution element} (RESEL), so we have far less number of effective $p$ \cite{worsley.1992}. Smoothing also reduces the effect of image registration errors and high frequency noise. Gaussian kernel smoothing  introduces continuous hierarchical structure through scale space \cite{worsley.1996.HBM}. However, small-$n$ large-$p$ problems become critical in brain network modeling, where we need to correlate different voxels. In the small-$n$ large-$p$ setting, the sample covariance and correlation matrices are no longer positive definite. Subsequently, up to $p-n$ nodes are statistically dependent although there might be {\em no} true dependency at all. Thus, there is need to constrain the  the covariance or correlation matrices by regularization methods such as sparse network models. Unfortunately, for large $p$, many sparse models have severe computational bottlenecks \cite{chung.2015.TMI}. 

\begin{figure}[t]
\centering
\includegraphics[width=1\linewidth]{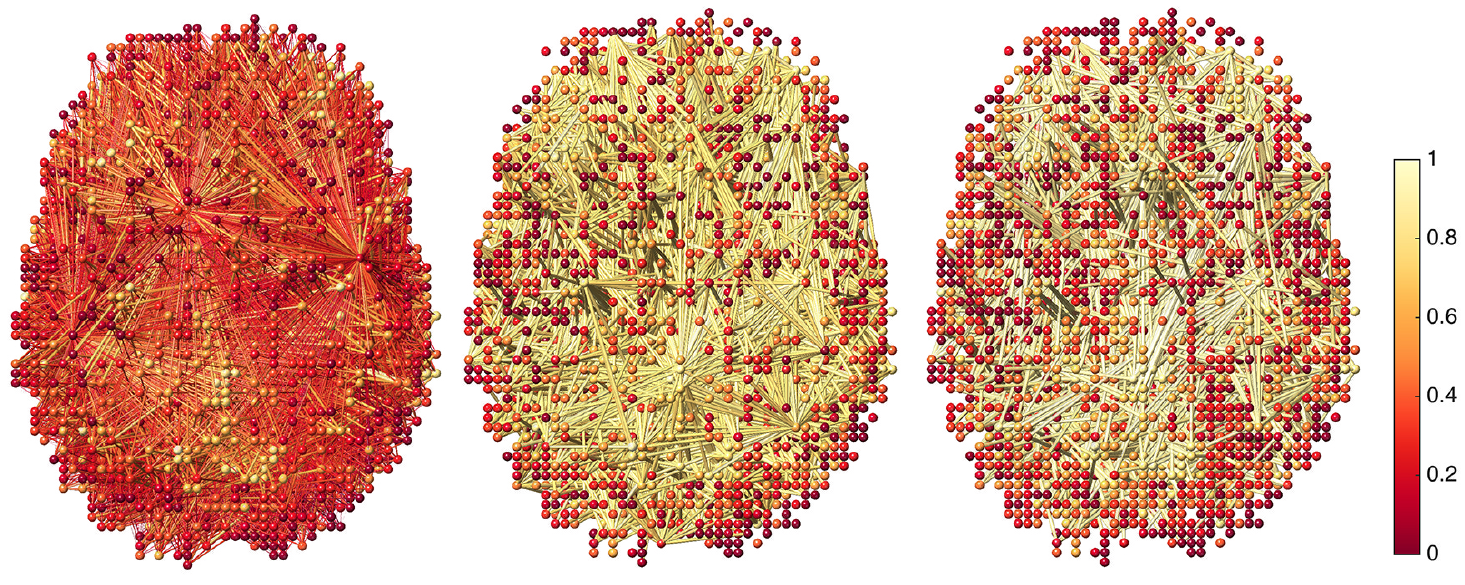}
\caption{Left: Dense fMRI correlation network consisting of more than 25000 nodes \cite{chung.2017.IPMI}. The network is so dense, simply displaying all the nodes and edges of the network is not very informative. It is necessary to represent such dense network more sparsely. The sparse correlation network model with sparse parameters $\lambda=0.7$ (middle) and $\lambda= 0.8$ (right). It can be shown that they form a nested hierarchy called the graph filtration.}
\label{fig:dense-network}
\end{figure}

There begin to emerge large-scale brain networks with more than 25000 nodes, where each voxel is taken as a network node (Figure \ref{fig:dense-network}) \cite{chung.2017.IPMI,eguiluz.2005,hagmann.2007,taylor.2017}. The size of such large-scale brain networks can easily match publicly available network data such as Stanford Large Network Dataset (\href{http://snap.stanford.edu/data}{\tt snap.stanford.edu/data}). In such large-scale networks, the small-$n$ large-$p$ problem will be more severe.  
%This type of big  data requires more {\em scalable} and {\em robust} solutions.  
%possibly using sparse penalties. 
%The following are the desirable properties of efficient methods for big data. 
%1) An effective method should require only a small subset of the full data in accurately estimating the underlying model. 
%2) An effective method should be unbiased and fast. 

%Many likelihood or L1-norm optimization techniques that perform well for reasonable sample size do not scale well with big imaging data.

%Optimization-based methods are very popular particularly in statistics and machine learning and relatively easy to implement when the cost functions are known. 
%Methods should likely to scale at least polynomially in rum time to even able to compute in tolerable time limit. Any method that scales exponentially with increased sample size is not going to be useful.
%Any method that requires the complete data set as the initial input is not going to be useful as well. 

\section{Large-scale brain networks}

Purely data-driven approaches for large-scale brain networks are not going to be computationally efficient or effective. It is often necessary to incorporate the first-order principles of brain networks into models to possibly reduce computational bottlenecks.

\subsection{Sparsity}
\begin{figure}[t]
\centering
\includegraphics[width=1\linewidth]{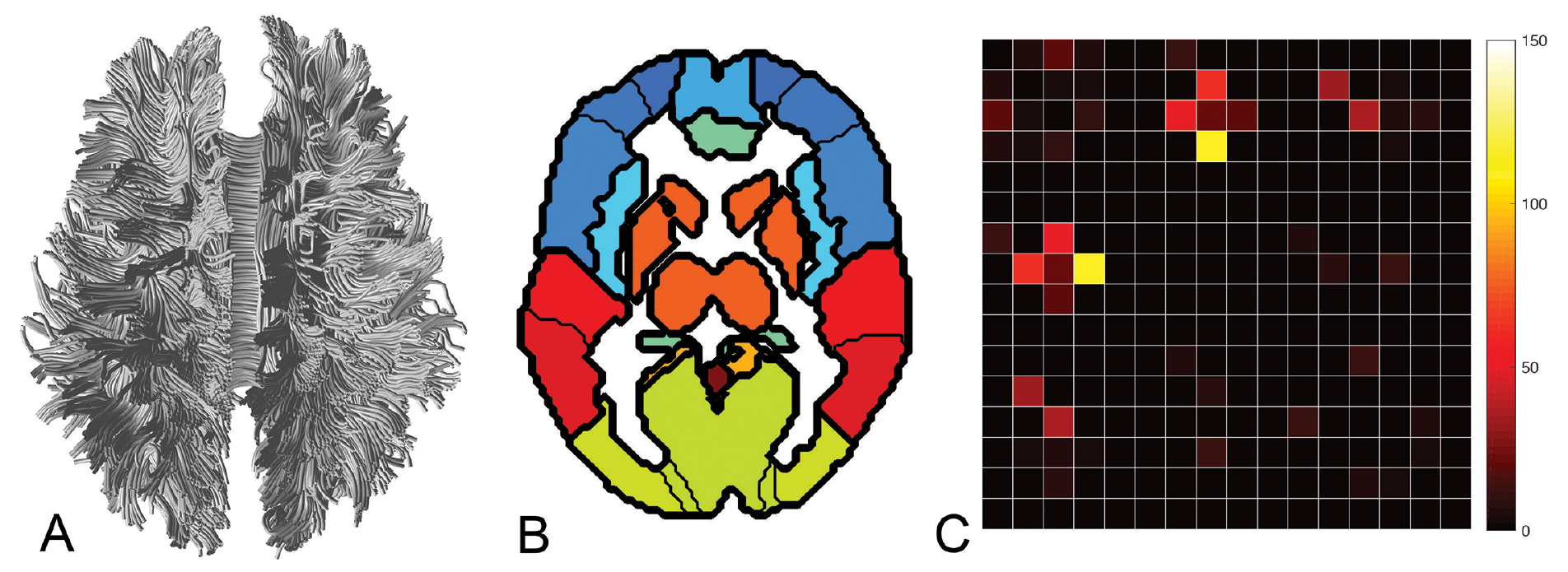}
\caption{A. White matter fiber tracts obtained from a tractography algorithm. B. The brain is parcellated into 116 disjoint regions. C. Connectivity matrix showing how each region is connected to other regions. Even thoroughly  fiber tracts are very dense, the resulting connective matrix is always sparse since not every part of brain is connected to each other \cite{chung.2017.BC}.}
\label{fig:sparse}
\end{figure}

\begin{figure}[t]
\centering
\includegraphics[width=1\linewidth]{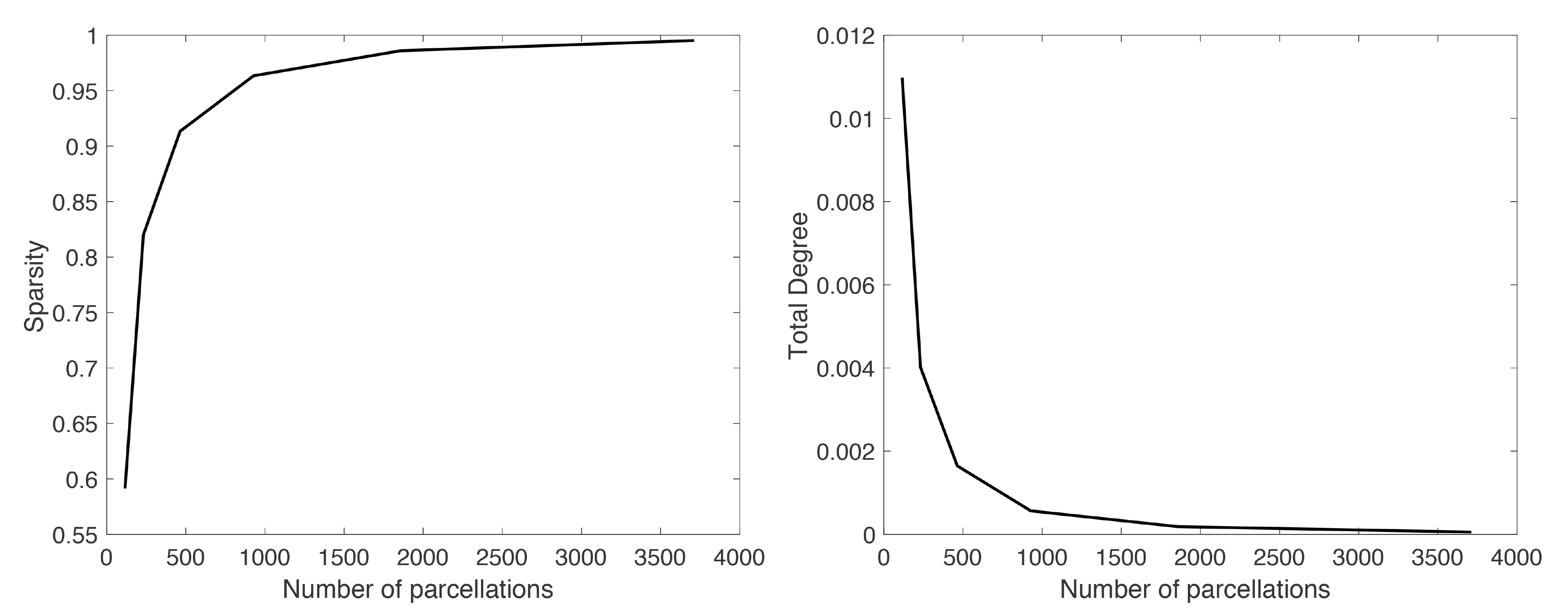}
\caption{Left: plot of sparsity over the number of  pacellations. The sparsity is measures as the ratio of zero entries over all entries in the connectivity matrix. Right: plot of total degree of nodes over the number of pacellations. The vertical axis measures the ratio of the total number of connections over every possible connection. The plots all show the sparse nature of brain networks at any spatial scale.}
\label{fig:sparseplot}
\end{figure}

At the microscopic level, the activation of cortical neurons in the brain show {\em sparse} and widely distributed patterns \cite{histed.2009}. At the macroscopic level, diffusion tensor imaging (DTI) can produce up to a half million white matter fiber tracts per brain. Even then not every part of the brain is anatomically connected to other parts of the brain but sparsely connected \cite{chung.2017.BC}. This can be seen from Figure \ref{fig:sparse}, where the brain is parcellated into 116 disjoint regions and the number of white matter fiber tracts passing between the regions is used in constructing the structural connectivity matrix \cite{chung.2017.BC}. Even though the white matter fibers are very dense, the resulting connectivity matrix is sparse. For $ 116 \times 116$ connectivity matrix, 60\% of entries are zeros. As we increases the number of parcellations, the sparsity increases while the the total degree of  all nodes decreases (Figure \ref{fig:sparseplot}). Note the degree of nodes counts the number of connections at a node. Thus, it also measures the sparsity of the network.

In fMRI studies, functional connectivity, which measures the dependency of brain activity in one region to another region, is often measured by correlation, covariance or spectral coherence  of fMRI time time series. Since the brain does not  activate everywhere simultaneously \cite{chung.2017.IPMI}, functional connectivity is also expected to be not dense but sparsely clustered.  It is reasonable to assume both functional and anatomical brain networks are sparsely connected at the both microscopic and macroscopic levels. Thus, there is strong biological justifications for modeling brain networks sparsely. 

%In brain imaging, there have been many  attempts to identify high-dimensional imaging features via multivariate approaches including network features \cite{worsley.2005.neural}. %,lerch.2006,he.2007,rao.2008,cao.1999.correlation,he.2008}. 
The small-$n$ large-$p$ problem in brain imaging often produces under-determined models with infinitely many possible solutions. Such problems are usually remedied by regularizing  the systems with additional sparse penalties. Sparse models used in brain imaging include compressed sensing  (CS) \cite{lee.2011.TMI}, sparse correlations \cite{chung.2017.IPMI}, least absolute shrinkage and selection operator (LASSO) \cite{huang.2009,lee.2011.TMI}, sparse canonical correlations \cite{avants.2010} and graphical-LASSO \cite{chung.2015.TMI,huang.2009}. Most of these sparse models require optimizing $L1$-norm penalties, which has been the major computational bottleneck for solving large-scale problems in brain imaging. Thus, almost all sparse brain network models have been restricted to a few hundreds nodes or less. 2527 MRI features used in {a LASSO model} for Alzheimer's disease \cite{xin.2015} is probably the largest number of features used in any sparse model {in the brain imaging literature}.  Recently, a more scalable large-scale sparse brain network models, where each voxel is a network node, %($p >$ 25000) 
are begin to emerge \cite{chung.2017.IPMI}. For such large-scale network construction, faster scalable algorithms are needed. In \cite{chung.2017.IPMI}, the computational bottleneck of $L1$-optimization is overcame by simplifying the sparse network problem into an orthogonal design. Other promising methods include a constrained $L_1$-minimization estimator (CLIME) \cite{wang.2016} and faster computations for graphical-LASSO \cite{witten.2011} although they were never applied to large-scale brain networks yet. 

%There are few previous studies at speeding up the computation for sparse models. By identifying block diagonal structures in the estimated (inverse) covariance matrix, it is possible to reduce the computational burden in the penalized log-likelihood method \cite{mazumder.2012}.  
%Specifically, we propose a novel sparse network model based on {\em cross-correlations}. Although cross-correlations are often used in sciences in connection to times series and stochastic processes \cite{worsley.2005.neural,worsley.2005.royal}, the sparse version of cross-correlation has been somewhat neglected. \\
%However, we can show that our method is made into related to the orthogonal design in 
%LASSO (least absolute shrinkage and selection operator) can be done without numerical optimization if the design matrix is orthogonal \cite{tibshirani.1996}.
%The proposed method also differs from \cite{tibshirani.1996} in that our problem is not exactly orthogonal. 

\subsection{Hierarchy}
Brain networks are fundamentally {\em multiscale}. Intuitive and palatable biological hypothesis is that brain networks are organized into {\em hierarchies} \cite{betzel.2016}. A brain network at any particular sale might be subdivided into subnetworks, which can be further subdivided into smaller subnetworks in an iterative fashion. There have been various attempts at modeling brain networks at multiple scales \cite{betzel.2016,chung.2015.TMI,chung.2017.IPMI,lee.2012.TMI} . Unfortunately, many multiscale models give raise to conflicting topological structures of the networks from one scale to the next.  For instance, the estimated modular structure in the multiscale community detection problem usually do not have continuity over different resolution parameters \cite{betzel.2016}.

Any sparse brain network model  is usually parameterized by a tuning parameter  that controls the sparsity of the solution. Increasing the sparse parameter makes the  solution more sparse. Thus, sparse models are inherently  multiscale, where the scale of the model is determined by the sparsity. Many existing sparse network models use a fixed parameter $\lambda$ that may not be optimal in other datasets or studies. Depending on the choice of the sparse parameter, the final network structure will be different \cite{chung.2015.TMI,lee.2012.TMI}.  There is a need to develop a multiscale sparse network model that provide a consistent analysis results and interpretation regardless of the choice of parameter \cite{chung.2015.TMI,chung.2017.IPMI}.

{\em Persistent homology} may offer an effective framework in addressing the topological inconsistency in multiscale models. Instead of studying images and networks at a fixed scale, as usually done in traditional approaches, persistent homology summarizes the changes of topological features over different scales and identifies the most persistent topological features that are robust under  different scales. This robust performance under different scales is needed for network models that are parameter and scale dependent. Instead of building networks at one fixed parameter that may not be optimal, persistent homological approaches exploit the topological structure of the data and models. In doing so,  topologically consistent nested hierarchical networks called the {\em graph filtration} is obtained \cite{lee.2012.TMI,chung.2015.TMI}. Such a nested hierarchical structure can  further speed up various computations for even for large-scale networks with a billions of connections \cite{chung.2017.IPMI}.

%The  persistent homological brain network,  offers one such solution. By exploiting the hidden topological structures of an intractably complex network problem, we have recently shown that it is possible to reduce exponential run time to quadratic run time \cite{chung.2017.IPMI}. 

\section{Discussion}

We have presented two main characterizations (sparsity and hierarchy) of brain networks that should be utilized even in big data environments. We have further explored various statistical challenges related to such characterizations.

The issue of sparsity and hierarchy is highly relevant to other types of big network data such as social networks \cite{christakis.2007}, World Wide Web (WWW) \cite{adamic.1999} and genomic regulatory networks \cite{luscombe.2004}. Given any type of real world network, it is unlikely that all the nodes are densely connected to each other. It is expected that the network to have sufficient sparsity. Many large scale networks such as social networks and WWW show scale-free characteristic, which is the main characteristic of hierarchical networks. Although we don't expect all networks to be hierarchical or sparse, these aspect of brain network should be applicable to other big network data.

In terms of computation, many existing brain image analysis software such as SPM (\href{http:/www.fil.ion.ucl.ac.uk/spm}{\tt www.fil.ion.ucl.ac.uk/spm}) and AFNI (\href{http://afni.nimh.nih.gov}{\tt afni.nimh.nih.gov}) are not effective for big data. The general statistical premise of such mainstream tools is that all the image measurements are available in the computer memory and statistics are computed using all the data. However, in the big data setting, it may not be possible to fit all of the imaging data in a computer's memory, making it necessary to perform the analysis by adding one image at a time in a sequential manner. We need a way to incrementally update the statistical analysis results without repeatedly running the entire analysis whenever new images or parts of images are added.

An {\em online algorithm} is one that processes its inputted data in a sequential manner \cite{chung.2017.MICCAI}. Instead of processing the entire set of imaging data from the start, an online algorithm processes one image at a time. That way, we can bypass the memory requirement, reduce numerical instability and increase computational efficiency. With the ever-increasing amount of large-scale brain imaging datasets such as ADNI and HCP, the development of various online statistical method is warranted \cite{chung.2017.MICCAI}. Thus, here is an immediate need to develop the online version of  sparse or hierarchical network models although there are no such available methods yet. Even large-scale Pearson correlation coefficients can be computed using an online algorithm.

Existing statistical analysis packages such as MATLAB and R also assume all measurements to be available in the computer memory. Unless substantial modification to existing codes is made, we cannot even compute $t$-statistics for extremely large data that will not fit into computer memory using the built-in functions. Thus, there is a strong need to develop online algorithms for big data beyond brain imaging.

%Big imaging data often requires {\em fast unbiased approximate solutions}. The {\em iterative residual fitting} algorithm, originally developed for estimating more than 20000 spherical harmonic coefficients per brain, may offer one such solution \cite{chung.2007.TMI}. The algorithm sequentially breaks down one gigantic problem into smaller problems in an iterative fashion. 

\bibliographystyle{plain}
\bibliography{reference.2017.12.20}
\end{document}